\documentclass[prl,twocolumn,nofootinbib, preprintnumbers, superscriptaddress]{revtex4-1}

\usepackage{amsmath,amssymb,amscd,simplewick}
\usepackage{listings}
\usepackage{dsfont}
\usepackage{slashed}
\usepackage{color}
\usepackage{ulem}
\usepackage{float}

\usepackage{graphicx}
\usepackage{epstopdf}
\usepackage{subfigure}
\usepackage{epsfig}

\usepackage{xcolor}
\usepackage[colorlinks=true,
            linkcolor=blue,
            urlcolor=blue,
            citecolor=green,          
            bookmarks=true,
            bookmarksnumbered=true,
            breaklinks=true,
            pdfpagemode=Fullscreen,
            pdfstartview=FitBH]{hyperref}

\hypersetup{pdfauthor = {Shao-Feng Ge},
	     pdftitle = {}, 
	     pdfsubject = {}, 
             pdfkeywords = {}, 
	     pdfcreator = {LaTeX with hyperref package},
	     pdfproducer = {dvips + ps2pdf} }

\definecolor{gesfpurple}{rgb}{0.47,0.19,0.42}

\definecolor{gesflanse}{rgb}{0.00,0.50,0.50}

\definecolor{gesfblue}{rgb}{0.08,0.42,0.76}

\definecolor{gesfred}{rgb}{1,0,0}

\definecolor{gesfwhite}{rgb}{1,1,1}

\definecolor{gesfblack}{rgb}{0,0,0}

\newcommand{\geqn}[1]{\hypersetup{linkcolor=blue}(\ref{#1})\hypersetup{linkcolor=blue}}
\newcommand{\gfig}[1]{{\hypersetup{linkcolor=violet}Fig.~\ref{#1}\hypersetup{linkcolor=blue}}}

\newcommand{\fp}{\slashed p}

\graphicspath{{figs/}}

\begin{document}

\title{Solar Neutrino Scattering with Electron into Massive Sterile Neutrino}
\author{Shao-Feng Ge}
\email{gesf@sjtu.edu.cn}
\affiliation{Tsung-Dao Lee Institute \& School of Physics and Astronomy, Shanghai Jiao Tong University, China}
\author{Pedro Pasquini}
\email{ppasquini@sjtu.edu.cn}
\affiliation{Tsung-Dao Lee Institute \& School of Physics and Astronomy, Shanghai Jiao Tong University, China}
\author{Jie Sheng}
\email{shengjie.physics@gmail.com}
\affiliation{Tsung-Dao Lee Institute \& School of Physics and Astronomy, Shanghai Jiao Tong University, China}
\affiliation{Physics Department, Jilin University, Changchun, Jilin Province, China}

\begin{abstract}
The recent Xenon1T excess can be explained by solar
neutrino scattering with electron via a light mediator,
either scalar or vector, in addition to many other
explanations from the dark sector. Since only the
recoil electron is observable, a keV sterile neutrino
instead of an active neutrino can appear in the final
state. The sterile neutrino allows pseudoscalar
mediator to explain the Xenon1T excess which was
thought impossible. In addition, nonzero recoil energy
lower bound arises from the sterile neutrino mass,
which can be used to testify if the
sterile neutrino is massive or not.
We also briefly discuss the case of sterile
neutrino final state with light $Z'$ mediator.
\end{abstract}

\maketitle 

\begin{center}
{\bf Introduction}
\end{center}

The recent Xenon1T data contains a low-energy peak excess in
the electron recoil spectrum \cite{Aprile:2020tmw}.
Although it is possible to explain this peak excess
with residual tritium background 
\cite{Aprile:2020tmw,Robinson:2020gfu}, more
data or independent measurement is needed before
making a decisive conclusion. It is too soon to call
this Xenon1T signal an anomaly. On the other hand,
this low energy peak can also be explained with
new physics beyond the Standard Model (SM).

The largest category is dark sector which contains
several possibilities. To explain the observed
low-energy excess with dark matter (DM),
\cite{Kannike:2020agf} pointed out that the DM has
to move faster, $v \gtrsim 0.05$, than the
conventional non-relativistic DM confined in our
galaxy, which typically has velocity
$v \sim 10^{-3}$ ($\mathcal O(100\,\mbox{km/s}$).
This is followed by many attempts in various
scenarios, including boosted DM \cite{Fornal:2020npv},
cosmic-ray boosted DM \cite{Giudice:2017zke,Cao:2020bwd,Jho:2020sku},
inelastic cosmic-ray boosted DM \cite{Su:2020zny},
warm DM decay \cite{Choi:2020udy},
DM heated in the Sun \cite{Chen:2020gcl},
inelastic DM-electron scattering from heavy to light
\cite{Harigaya:2020ckz,Lee:2020wmh},
inelastic DM-electron scattering from light to heavy
\cite{Bell:2020bes},
$2 \rightarrow 4$ annihilating DM \cite{Du:2020ybt}, $3\rightarrow2$ Co-SIMP DM~\cite{Smirnov:2020zwf},
shinning DM emitting photon to fake electron recoil
with Rayleigh operator \cite{Paz:2020pbc},
electromagnetic de-excitation \cite{Baryakhtar:2020rwy},
Migdal effect with electron signal from nuclei recoil
\cite{Dey:2020sai}, mirror DM \cite{Zu:2020idx}, and Hydrogen atom decay~\cite{McKeen:2020vpf}.
There is also some discussion about using collider
search to supplement the direct detection
\cite{Primulando:2020rdk}.
Although the solar axion
explanation is in tension with astrophysical
constraints \cite{Aprile:2020tmw,DiLuzio:2020jjp},
reexamination shows new possibility
\cite{Gao:2020wer, Bloch:2020uzh, DeRocco:2020xdt,Dent:2020jhf} and
axion-like-particle (ALP) dark matter is much
less constrained \cite{Takahashi:2020bpq,Buch:2020mrg,
Bloch:2020uzh}. The light boson category  also
includes relaxion \cite{Budnik:2020nwz} and
dark photon
\cite{Alonso-Alvarez:2020cdv,Nakayama:2020ikz,An:2020bxd,Bramante:2020zos}.

The electron recoil by solar neutrinos is also a
major category. Large magnetic momentum from the Majorana nature of neutrinos~\cite{Bell:2005kz,Bell:2006wi} has also being used to explain the Xenon1T excess~\cite{Aprile:2020tmw,Khan:2020vaf,Chala:2020pbn}. In addition to the photon mediator,
either a light scalar \cite{Boehm:2020ltd,Khan:2020vaf}
or $Z'$
\cite{Boehm:2020ltd,Khan:2020vaf,Bally:2020yid,AristizabalSierra:2020edu,Lindner:2020kko} is needed.
Although the neutrino charge radius
\cite{Khan:2020vaf} and muon magnetic momentum
\cite{Amaral:2020tga} was additionally studied,
they can only provide a quite flat recoil spectrum
and hence cannot explain the Xenon1T excess.
It is a common feature of having light mediators
for the solar neutrino explanations. Both scalar
and vector mediators have been discussed. And it
is explicitly claimed that the pseudoscalar mediator
does not work \cite{Boehm:2020ltd} due to lack of
low recoil enhancement \cite{Budnik:2020nwz,Boehm:2020ltd}, which is not
necessarily true.

We propose a new possibility with sterile neutrino
in the final state, in addition to the light 
mediator. It allows pseudoscalar to provide
a $1/T_r$ enhancement at low energy and hence can
explain the Xenon1T excess. In addition,
the finite mass of the sterile neutrino leads to
a sharp cutoff on the lower side of the electron
recoil spectrum for fixed neutrino energy.
This provides the DM direct detection experiments
a chance of not just probing the existence of DM but also measuring the companion particle mass.
The same scenario also applies for a light $Z'$
mediator as we briefly discuss at end.

\vspace{3mm}
\begin{center}
{\bf Sterile Neutrino and Light Mediator}
\end{center}

Let us consider a light scalar mediator with both
neutrino and electron,
\begin{equation}
  \mathcal L_{\rm int}
=
  \bar \nu (y^\nu_S + \gamma_5 y^\nu_P) \phi \nu_s
+ \bar e (y^e_S + \gamma_5 y^e_P) e \phi
+ h.c.,
\end{equation}
with both scalar and pseudoscalar couplings to
keep general. This kind of coupling can arise from
the mixing of scalar $\phi$ with the SM Higgs.
The Yukawa term requires a right-handed neutrino
that has no SM gauge interactions and hence is a
sterile neutrino. It is possible for this sterile
neutrino to obtain a Majorana mass term. The
sterile neutrino $\nu_s$ has mass $m_s$
$\sim O(100\,{\rm keV})$ and the scalar mediator
$\phi$ has mass $m_\phi \lesssim 30\,\mbox{keV}$.

When scattering with electron, the light mediator
would introduce a $1/(q^2 - m^2_\phi)$ propagator.
If the mediator is light enough,
$m^2_\phi \ll q^2 = 2 m_e T_r$, $1/T_r$ enhancement
at low energy naturally appears. The Xenon1T
peak excess, $T_r \approx (2 \sim 3)\,\mbox{keV}$,
corresponds to $m_\phi \ll (40\sim 60)\,\mbox{keV}$.
Nevertheless, whether
the differential cross section has such feature or
not is still subject to the scattering matrix 
element $\overline{|\mathcal M|^2}$ \cite{FeynCalc},
\begin{equation}
\hspace{-2mm}
  4 m_e (2 m_e T_r + m^2_s)
  \frac {(y^\nu_S y^e_S)^2 (2 m_e + T_r) + (y^\nu_P y^e_P)^2 T_r}
        {(2 m_e T_r + m^2_\phi)^2}.
\label{eq:M2}
\end{equation}
The first term in the numerator is the scalar
contribution while the second one comes from the pseudoscalar coupling.
The difference formally emerges from the electron
spinor trace.
\begin{subequations}
\begin{eqnarray}
  \mbox{Scalar}
& : &
  |\mathcal M|^2
\propto
  \mbox{Tr} \left[ (\fp_e + m_e) (\fp'_e + m_e) \right],
\label{eq:trace-scalar}
\\
  \mbox{Pseudoscalar}
& : &
  |\mathcal M|^2
\propto
  \mbox{Tr}
\left[ (\fp_e + m_e) \gamma_5 (\fp'_e + m_e) \gamma_5 \right].
\qquad
\label{eq:trace-pseudoscalar}
\end{eqnarray}
\end{subequations}
Due to the presence of $\gamma_5$, the scalar
matrix element \geqn{eq:trace-scalar}
becomes $4 (m^2_e + p_e \cdot p'_e)$ while the pseudoscalar
one is $4(m^2_e - p_e \cdot p'_e)$ instead. With
$p_e \cdot p'_e = m_e (T_r + m_e)$. Hence, the scalar
and pseudoscalar terms become $4 m_e (2 m_e + T_r)$ 
and $4 m_e T_r$, respectively. The pseudoscalar
terms scales linearly with $T_r$ while the scalar
one is almost flat.

The scalar and pseudoscalar couplings contribute
separately. To see the features of this new
interaction in \geqn{eq:M2} more clearly, let
us first omit the mediator mass $m_\phi$ and take
$2 m_e \gg T_r$ into consideration. For a light
sterile neutrino, $2 m_e T_r \gg m^2_s$, the scalar
term has $1/T_r$ peak while the pseudoscalar one
becomes almost independent of the electron recoil
energy. For light final state such as the active
neutrinos in the SM, the pseudoscalar mediator can
not explain the Xenon1T excess
\cite{Cerdeno:2016sfi,Boehm:2020ltd}.

\begin{figure}[t]
\centering
\includegraphics[width=0.45\textwidth,height=56mm]{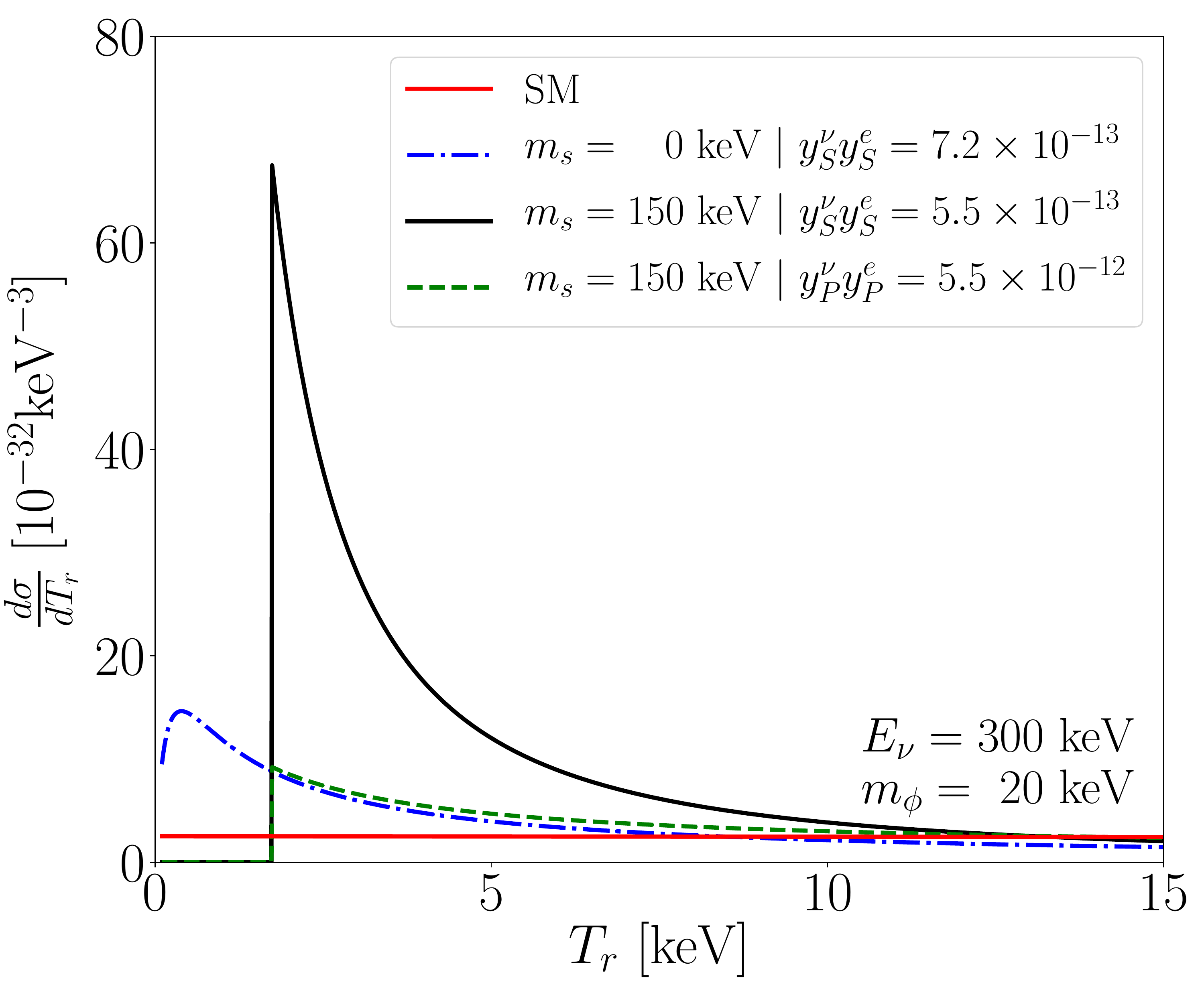}
\caption{The electron recoil energy spectrum
for the SM background and sterile neutrino
final state with scalar mediator.}
\label{fig:dSigma}
\end{figure}

Nevertheless, this conclusion is not necessarily
true in the presence of final-state sterile neutrino.
For $m^2_s \gtrsim 2 m_e T_r$, the prefactor
$2 m_e T_r + m^2_s \approx m^2_s$ in \geqn{eq:M2}
no longer has linear dependence on the electron
recoil energy $T_r$ but becomes almost flat.
This introduces $1/T_r$ to the pseudoscalar
contribution and enhance the scalar one to $1/T^2_r$.
In addition to the energy recoil peak introduced
by a light mediator that one usually expects, a
massive final state can do the same thing.
The scenario of hidden neutrino in the
final state \cite{Bally:2020yid} with mass at
sub-eV scale and a $Z'$ mediator is quite
different from the one we consider here.


From the scattering matrix element
$\overline{|\mathcal M|^2}$ to the differential
cross section $d \sigma / d T_r$,
\begin{equation}
  (2 m_e T_r + m^2_s)
  \frac {(y^\nu_S y^e_S)^2 (2 m_e + T_r) + (y^\nu_P y^e_P)^2 T_r}
        {8\pi E_\nu^2 (2 m_e T_r + m^2_\phi)^2},
\end{equation}
no extra $T_r$ is introduced.
We show the electron recoil energy spectrum in
\gfig{fig:dSigma} which clearly demonstrates
a surging
peak at low energy. The SM contribution is mainly
from the heavy $Z$ and $W$ mediators with flat
recoil spectrum \cite{Giunti:2007ry}. The blue
curve for scalar and massless neutrino in the
final state roughly overlaps with the dashed curve
for pseudoscalar curve with $m_s = 100\,\mbox{keV}$
due to the $1/T_r$ dependence as we elaborated
above. If the final-state neutrino is also massive,
the scalar contribution receives one more $1/T_r$
enhancement and leads to the black curve in
\gfig{fig:dSigma}, which shows
clearly the effect of a massive sterile neutrino
in explaining the low-energy recoil signal observed
by Xenon1T.

The sterile neutrino as DM can also introduce
a peak in the low energy recoil spectrum 
\cite{Campos:2016gjh}. The mass of the sterile
neutrino in the initial
state is converted to kinetic energy of the
final-state particles. With sterile neutrino mass,
$m_s\lesssim 40 $ keV, the electron recoil energy
receives a natural upper limit at the keV scale
and hence a peak. In this scenario, there is no
need to involve a light mediator and the SM
$Z$ boson mediation is enough to explain the low
energy peak.

\begin{figure}[t]
\centering
\includegraphics[width=0.42\textwidth]{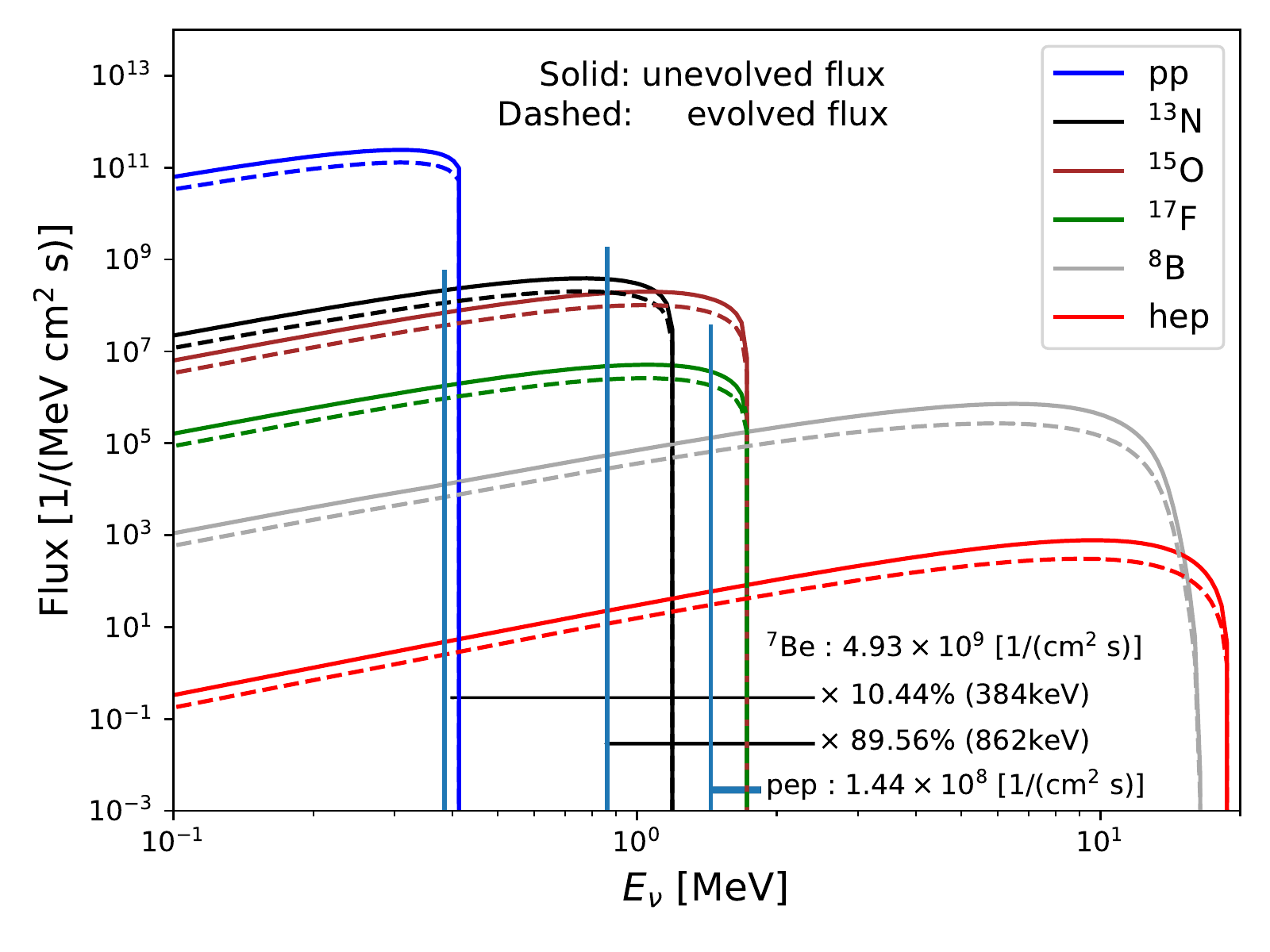}
\includegraphics[width=0.42\textwidth]{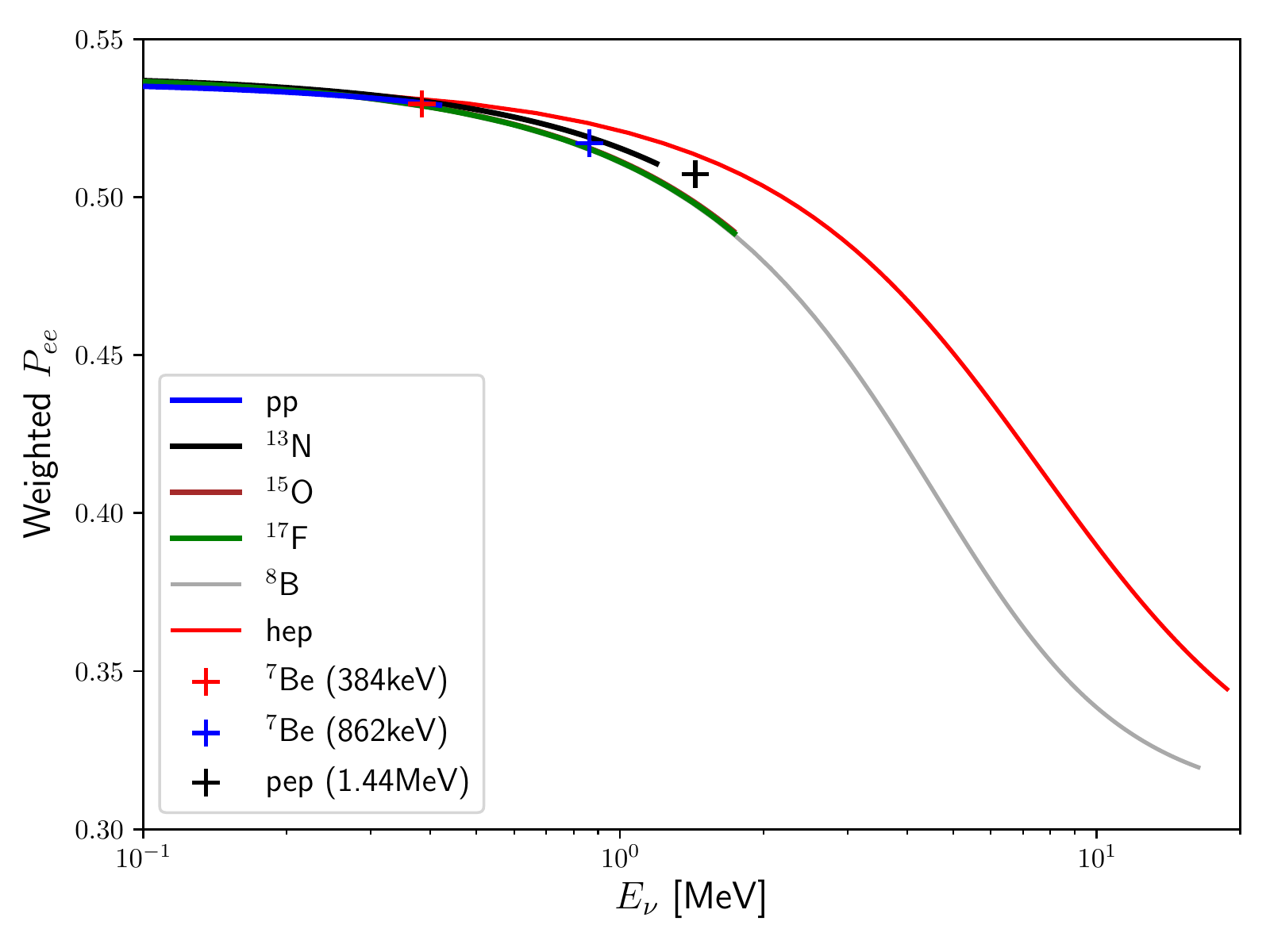}
\caption{The solar neutrino fluxes and the
transition probabilities.}
\label{fig:nuSolar}
\end{figure}

\vspace{3mm}
\begin{center}
{\bf Solar Neutrino}
\end{center}

We used the NuPro package \cite{NuPro} to simulate the
evolution \cite{Wolfenstein:1977ue,resonant}
of solar neutrinos
\cite{Maltoni:2015kca,Wurm:2017cmm,Vissani:2017dto}.
The solar neutrinos are produced from the $pp$ chain
and CNO cycle nuclear reactions according to which
the neutrino fluxes can be predicted
\cite{Vinyoles:2016djt}. In total, there are 6
continuous fluxes ($pp$, $^{13}$N, $^{16}$O, $^{17}$N,
$^8$B, and $hep$) and 3 monoenergetic ones
(the $^7$B flux has 10.44\% at 384\,keV and 89.56\%
at 862\,keV, as well as $pep$ at 1.44\,MeV). The
fluxes without flavor transition have been
illustrated with solid lines in
the upper panel of \gfig{fig:nuSolar}.
Depending on the solar matter density,
composition, and temperature, the production rate
varies inside the Sun. Due to the extremely high
density inside the Sun, solar neutrinos evolve
adiabatically when propagating out and the transition
probability $P_{ee}$ depends on the neutrino
production location. The solar neutrino fluxes that
arrive at detector on the Earth is then the one
convoluted with transition probabilities, illustrated
as dashed lines in \gfig{fig:nuSolar}.
These effects have been
properly taken into account in NuPro. The lower
pannel of
\gfig{fig:nuSolar} shows the transition 
probabilities for different fluxes.
Between different fluxes, the transition probability
varies a lot, especially for the high energy part.
In our simulation,
the neutrino parameters are assigned to the best-fit
values from the latest global fits
\cite{deSalas:2020pgw,Capozzi:2020qhw}.

\vspace{3mm}
\begin{center}
{\bf Xenon1T Electron Recoil Signal}
\end{center}

\begin{figure}[t]
\centering
\includegraphics[width=0.45\textwidth,height=56mm]{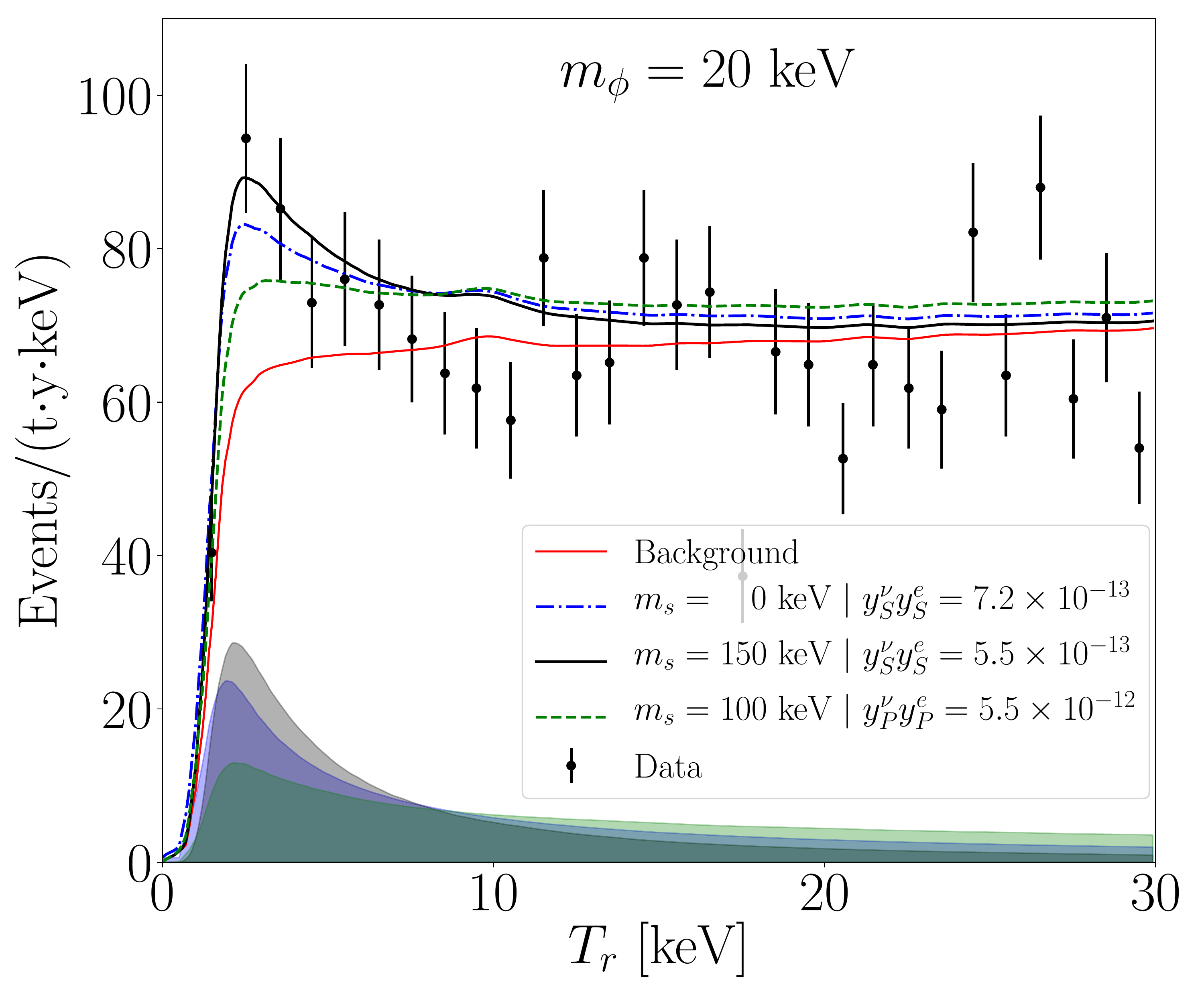}
\caption{The electron recoil event rate at
Xenon1T.}
\label{fig:eventRate}
\end{figure}

The \gfig{fig:eventRate} shows the signal event
rates for both scalar and pseudoscalar mediators.
The Xenon1T excess is observed in the Science Run 1
(SR1) data set with $0.65$ ton$\cdot$year exposure.
Although the recoil energy spectrum has a sharp
peak at low energy, we have to consider the finite
energy resolution and the
detection efficiency \cite{Aprile:2020tmw}.
The energy resolution can be parametrized as
$\sigma_{T_r} = a / \sqrt{T_r/{\rm keV}} + b$
with $a = 31.71 \pm 0.65$ and $b = 0.15 \pm 0.02$
\cite{Aprile:2020yad}. For simplicity, we just use
the central values of $a$ and $b$.
We assign $m_s = 150\,\mbox{keV}$ for the
scalar mediator and $m_s = 100\,\mbox{keV}$ for
the pseudoscalar. In both cases,
$m_\phi = 0\,\mbox{keV}$ can explain the Xenon1T 
signal.

It should be emphasized that the signal spectrum
in \gfig{fig:eventRate} drops
faster for a massive sterile neutrino than the
massless case. This is
because the recoil energy can receive a nonzero
lower limit for a massive sterile neutrino final
state as indicated in \gfig{fig:dSigma},
$T^-_r \leq T_r \leq T^+_r$ with,
\begin{eqnarray}
  T^\pm_r
& = &
  \frac 1 {2 s}
\left\{
  ( s + m^2_e - m^2_s ) E_0
\right.
\\
& \pm &
\left.
  |{\bf p}_0|
\sqrt{
\left[ s - (m_e + m_s)^2 \right]
\left[ s - (m_e - m_s)^2 \right]}
\right\}
- m_e.
\nonumber
\label{eq:Elimits}
\end{eqnarray}
The center-of-mass energy is,
$s = m_e (m_e + 2 E_\nu)$, $E_0 = E_\nu + m_e$,
and $|{\bf p}_0| = E_\nu$. For massless final-state
neutrino, $T^-_r = m_e$ and consequently $T_r \geq 0$.
In the presence of massive sterile neutrino final
state, there is no recoil signal below $T^-_r - m_e$.
It is then possible to experimentally justify if
the final state companion particle is massless or not.
From the gap size low recoil spectrum
shape, we may infer the companion particle
mass.
The electron recoil signal of DM direct detection
experiment can not only probe the existence of DM
but also the mass of the companion particle.
This is not easy at the conventional detectors
and new concept detection techniques may help.

\vspace{3mm}
\begin{center}
{\bf Interplay with the SM Counterparts}
\end{center}

In previous discussions, the SM and new physics
contributions are essentially independent of each
other. Here we try to compare these two contributions
by first comparing their sizes and then briefly
discuss the possible interference between them.

With the integration limits $T^\pm_r$ in
\geqn{eq:Elimits}, we can readily obtain the total
cross section,
\begin{eqnarray}
  \sigma
& = &
  \frac {(y^\nu_S y^e_S)^2}{16 \pi m_e E^2_\nu}
\left[
  2 m_e \log \frac {T^+_r}{T^-_r}
- m^2_s \left( \frac 1 {T^+_r} - \frac 1 {T^-_r} \right)
\right]
\nonumber
\\
& + &
  \frac {(y^\nu_P y^e_P)^2}{32 \pi m^2_e E^2_\nu}
\left[
  2 m_e (T^+_r - T^-_r)
+ m^2_s \log \frac {T^+_r}{T^-_r}
\right].
\end{eqnarray}
A basic feature is the integration limits $T^\pm_r$
scales with neutrino energy $E_\nu$ almost linearly
for large neutrino energy. Consequently, the total
cross section is suppressed for $E_\nu \gg m_e, m_s$
with a $1/E_\nu$ scaling.
This behavior is quite different from the SM counterpart
\cite{Giunti:2007ry} which increases linearly with
neutrino energy.

In earlier discussions,
the SM diagram with active neutrino $\nu_e$
in the final state does not interfere with the
sterile neutrino contribution due to different
final states, one is active neutrino and the other
sterile neutrino. This is based on the assumption
of ignoring the active-sterile mixing which is
always achievable since the existence of sterile
neutrino has not been established by neutrino
oscillation experiments
\cite{Giunti:2019aiy,Diaz:2019fwt,Boser:2019rta}, 
although the decay of a keV neutrino could explain
the LSND and MiniBooNE excess \cite{deGouvea:2019qre}.

One may imagine the situation that the active-sterile
mixing can introduce flavor-changing neutral
current in neutrino interactions. Then, even for
the SM interactions, both neutral and charged
currents, the sterile neutrino can appear in the
final state. For both cases, the scattering matrix
element with $Z/W$ mediator is suppressed by the
active-sterile mixing $\theta_{as}$,
$\mathcal M^{\rm s}_{Z/W} \approx
\theta_{as} \mathcal M^{\rm SM}_{Z/W}$.
The corresponding interference term is then of the
order $\theta_{as} \mathcal M^{\rm SM}_{Z/W}
\mathcal M_\phi$ where $\mathcal M_\phi$
is the scalar-mediated contribution.
As indicated in \gfig{fig:dSigma}, the
contribution of these two contributions is roughly
the same when explaining the Xenon1T excess,
$|\mathcal M^{\rm SM}_{Z/W}|^2 \sim |\mathcal M^{as}_\phi|^2$.
The interference term is then controlled by the
active-sterile mixing $\theta_{as}$ and can be
easily suppressed.

The sterile neutrino scenario has implementations
in both neutrino oscillation
\cite{Giunti:2019aiy,Diaz:2019fwt,Boser:2019rta}
and dark matter 
\cite{Adhikari:2016bei,Boyarsky:2018tvu}.
The keV sterile neutrino discussed in this paper
belongs to the dark matter category and hence is
largely unconstrained by the neutrino oscillation
experiments. The strongest constraint comes from
meson decay experiments and for electron neutrino
it is of order $|y^\nu_{S,P}| \lesssim 10^{-3}$ \cite{Pasquini:2015fjv,Berryman:2018ogk,Dror:2020fbh,deGouvea:2019qaz}. 
At the same time, Big Bang Nucleosynthesis 
requires that $|y^e_S| \lesssim 5 \times 10^{-10}$~\cite{Babu:2019iml} 
if the scalar mediator is kept in thermal 
equilibrium with the primordial plasma before $T \sim 1$ MeV, 
which would decrease the deuterium abundance. 
The combined bound is
$|y^\nu_S y^e_S| \lesssim 5 \times 10^{-13}$
which our parameter choice satisfies.

\begin{figure}[t]
\centering
\includegraphics[width=0.45\textwidth,height=56mm]{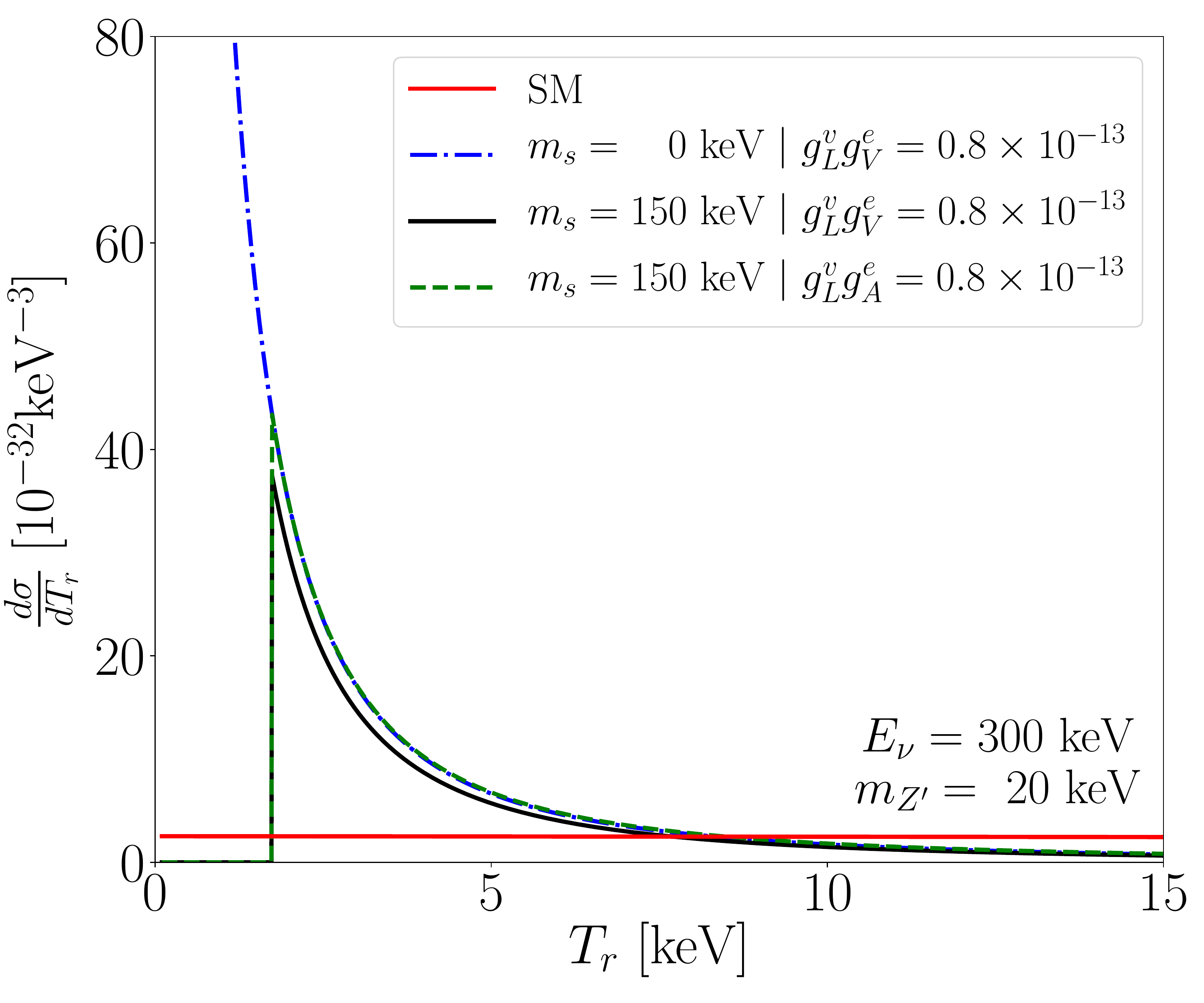}
\caption{The electron recoil energy spectrum
for the SM background and sterile neutrino
final state with scalar mediator.}
\label{fig:dSigma_vector}
\end{figure}
a
\vspace{3mm}
\begin{center}
{\bf Vector Mediator}
\end{center}

The same scenario of sterile neutrino in the final
state also works with a light vector boson mediator.
Since the recoil energy gap arises from kinematics,
its application is not limited to the scalar force.

For simplicity, we consider only the situation where
$Z'$ couples to the left-handed neutrinos,
\begin{eqnarray}
  \mathcal L_{\rm int}
=
  g^\nu_L \overline \nu \gamma^\mu P_L \nu_s Z'_\mu
+
  \overline e (g^e_V - g^e_A \gamma_5)\gamma^\mu e Z'_\mu.
\end{eqnarray}
On the other hand, the electron coupling can have
either vector or axial-vector current coupling
with $Z'$. The corresponding differential
cross section is,
\begin{eqnarray}
  \frac{d \sigma}{d T_r}
& = &
  \frac{(g^\nu_L g^e_{V,A})^2}
       {2\pi E_\nu^2}
\left[
  \frac{2m_e(2E_\nu^2+T_r^2-T_r(2 E_\nu \pm m_e))}
       {(2 T_r m_e + m_{Z'}^2)^2}
\right.
\nonumber 
\\
& &
\hspace{19mm}
\left.
-
  \frac{m_s^2 (2 E_\nu - T_r \pm m_e)}
       {(2T_rm_e+m_{Z'}^2)^2}
\right],
\label{eq:dSigma_vector}
\end{eqnarray}
where + ($-$) stands for vector (axial-vector)
current interaction with non-vanishing $g^e_{V,A}$
(vanishing $g^e_{A,V}$), respectively.
It is interesting to see that the combination
of neutrino and electron couplings
$(g^\nu_L g^e_{V,A})^2$ appear as overall factors. The
differential cross section has exactly the same
functional form no matter $Z'$ couples to electron
with vector or axial-vector interactions. This is
very different from the scalar versus pseudoscalar
comparison. Since the
neutrino energy $E_\nu$ is much larger than the
electron recoil energy $T_r$, the numerators are
insensitive to $T_r$ and the two terms of \geqn{eq:dSigma_vector} always have $1/T^2_r$ enhancement.

\begin{figure}[t]
\centering
\includegraphics[width=0.45\textwidth,height=56mm]{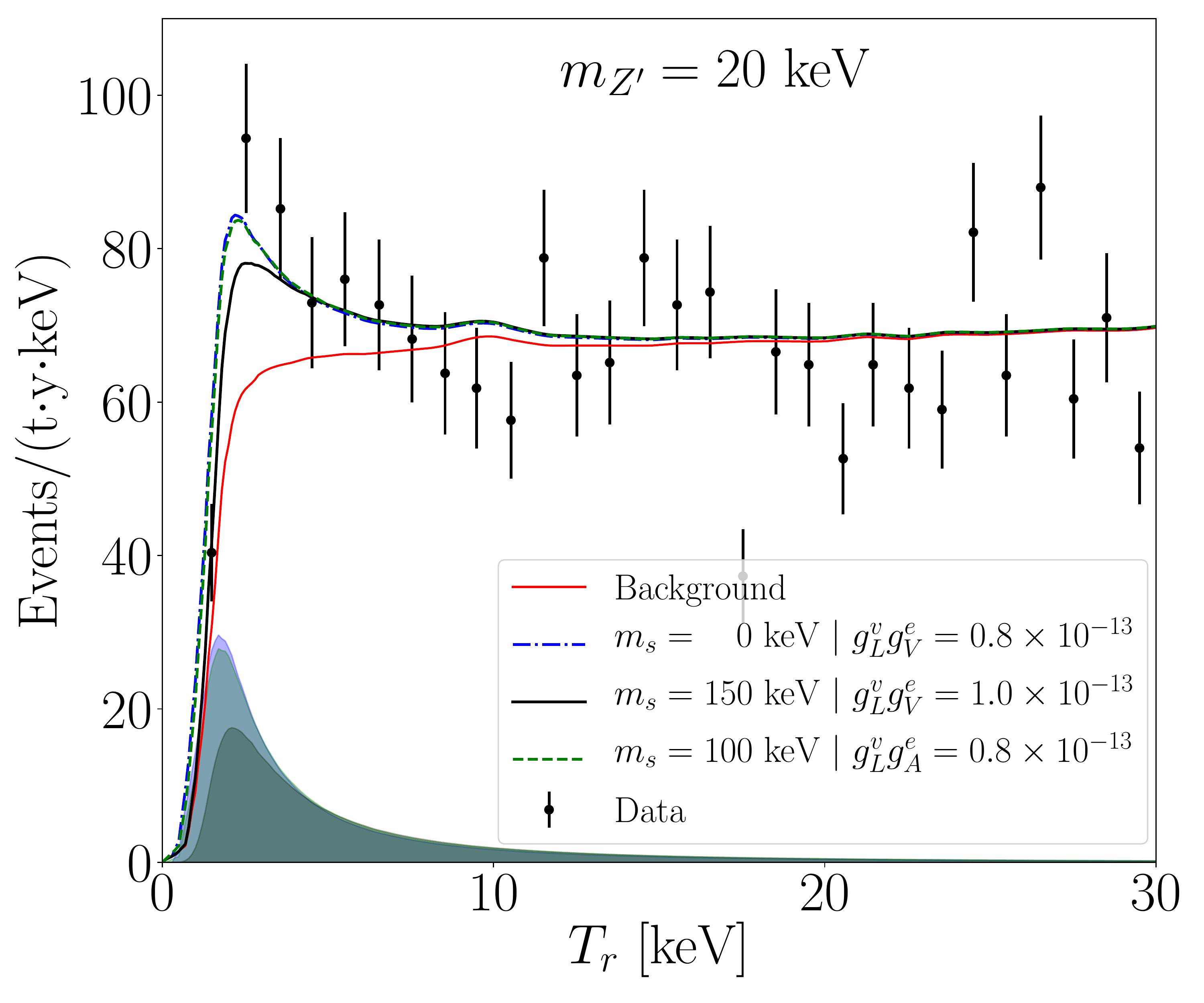}
\caption{The electron recoil event rate at
Xenon1T.}
\label{fig:eventRate_vectr}
\end{figure}

\gfig{fig:dSigma_vector} shows the differential
cross section as a function of the electron recoil
energy $T_r$. As expected, the sterile neutrino
mass $m_s$ places a much less significant role
than the scalar case, the two terms of
\geqn{eq:dSigma_vector} have roughly the same
contribution. This can be clearly seen as the
same shape of the three sterile neutrino curves.
The only major difference between massive and
massless sterile neutrinos is the suppressed spectrum
at the lower end for the former case.
This can lead to testable effect with better
resolution, lower threshold, and higher efficiency.
The event rate spectrum at Xenon1T is shown in
\gfig{fig:eventRate_vectr}. The combination of
massive sterile neutrino and light vector mediator
can also explain the Xenon1T excess. 

\vspace{3mm}
\begin{center}
{\bf Non-Standard Interactions}
\end{center}

With light mediator coupled to both neutrino
and electron, neutrinos can feel the matter effect
that arises from the forward scattering with
electron. This leads to nonstandard interactions
for $Z'$ mediator,
\begin{eqnarray}
  \epsilon^{e}_{es}
=
  \frac{g^{\nu}_{V,A} g^e_{V,A}}
       {4\sqrt{2}G_{\rm F} m^2_{Z'}}
\sim
  g^\nu_{V,A}
  g^e_{V,A}
\left(
  \frac{10^{8}}{m_{Z'} [\mbox{keV}]}
\right)^2 \,.
\end{eqnarray}
For $m_{Z'} = 1 \sim 20\,\mbox{keV}$ and
$g^\nu_{V,A} g^e_{V,A} = 10^{-13}$,
the size of the NSI parameter is
$\epsilon^e_{es} = 3.6 \sim 1500$
with $\epsilon^e_{es} \equiv
g^\nu_{V,A} g^e_{V,A} / (V_{cc} m^2_{Z'})$
and $V_{cc}$ denoting the matter potential
from $W$ exchange.
Previous works considered the diagonal elements
$\epsilon^e_{ss}$~\cite{Liao:2018mbg,Liao:2016reh}
or cosmological impacts of ultra-light
mediators~\cite{Farzan:2019yvo} but not the
active-sterile off-diagonal element that we
discuss here. For comparison, the current bound
on the NSI with only active neutrinos favors 
a small value $|\epsilon_{ee}^e|\lesssim 0.05$
\cite{Farzan:2017xzy}, but only for the part
after subtracting a common diagonal contribution.

For scalar mediator, the matter effect appears as
correction to the neutrino mass term
\cite{SpinFlip,Liao:2015rma,Ge:2018uhz}.
With $m_\phi = 20\,\mbox{keV}$, the scalar NSI
parameters $\eta_{es} = 4\times 10^{-17}$ 
($\delta M \equiv y^\nu_{S,P} y^e_{S,P}/m^2_\phi$
and $\eta_{es} \equiv \delta M / \sqrt{\Delta m^2_{31}}$) is quite small.
For a sizable vector NSI, the
correction $E_\nu V_{cc} \epsilon_{es}$ is
to the mass squared term $M_\nu M^\dagger_\nu$
while the matter potential for
scalar is $M_\nu + \eta_{es} \sqrt{\Delta m^2_{31}}$. For roughly same size of scalar and vector NSI,
$\epsilon_{es} V_{cc} \sim \eta_{es} \sqrt{\Delta m^2_{31}}$,
the relative correction is
$E_\nu \epsilon_{es} V_{cc} / M_\nu M^\dagger_\nu \gg \eta_{es} \sqrt{\Delta m^2_{31}} / M_\nu$
since the neutrino energy is much larger than
the neutrino mass, $E_\nu \gg M_\nu$.
Then, we can safely omit the scalar NSI in the
parameter space considered in this paper.

So both the scalar and vector mediator cases
are safe from the NSI constraints.

\vspace{3mm}
\begin{center}
{\bf Conclusions}
\end{center}

In contrary to the common expectation that
pseudoscalar mediator cannot leave significant
contribution to low recoil phenomena, we notice
a massive final-state companion fermion can
introduce $1/T_r$ enhancement even for pseudoscalar
mediator. Although we propose this sterile neutrino
option to explain the Xenon1T excess, the same
scenario also applies for other situations such as
a light relativistic DM scattering into a massive
invisible fermion. This significantly enlarges the
model spectrum.
The sub-keV threshold detector for
low recoil energy scan to testify the existence of
massive companion particle is an interesting
direction to pursue.

\section*{Acknowledgements}

SFG is grateful to the Double First Class start-up
fund (WF220442604) provided by Shanghai Jiao Tong
University.

\end{document}